\documentclass[a4paper,12pt]{article}
\usepackage{amsmath}
\usepackage{amssymb}
\usepackage{amsfonts}
\usepackage{graphics}
\usepackage{epsfig}
\usepackage{graphicx}
\usepackage{epic}
\usepackage{eepic}

\newcommand{\cellsize}{10}
\newlength{\cellsz} 
\newsavebox{\cell}
\sbox{\cell}{\begin{picture}(\cellsize,\cellsize)
\put(0,0){\line(1,0){\cellsize}}
\put(0,0){\line(0,1){\cellsize}}
\put(\cellsize,0){\line(0,1){\cellsize}}
\put(0,\cellsize){\line(1,0){\cellsize}}
\end{picture}}
\newcommand\cellify[1]{\def\thearg{#1}\def\nothing{}%
\ifx\thearg\nothing
\vrule width0pt height\cellsz depth0pt\else
\hbox to 0pt{\usebox{\cell} \hss}\fi%
\vbox to \cellsz{
\vss
\hbox to \cellsz{\hss$#1$\hss}
\vss}}
\newcommand\tableau[1]{\vtop{\let\\\cr
\baselineskip -16000pt \lineskiplimit 16000pt \lineskip 0pt
\ialign{&\cellify{##}\cr#1\crcr}}}

\begin{document}

\renewcommand{\theequation}{\thesection.\arabic{equation}}
\thispagestyle{empty}
\vspace*{-1.5cm}

\begin{center}
{\Large\bf Generalized Curvature and Ricci Tensors for a \\
 Higher Spin Potential\\ and the Trace Anomaly in External Higher \\Spin
Fields in $AdS_{4}$ Space}\\

\vspace{4cm}
{\large Ruben Manvelyan ${}^{\dag\ddag}$ and Werner R\"uhl
${}^{\dag}$}
\medskip

${}^{\dag}${\small\it Department of Physics\\ Erwin Schr\"odinger Stra\ss e \\
Technical University of Kaiserslautern, Postfach 3049}\\
{\small\it 67653
Kaiserslautern, Germany}\\
\medskip
${}^{\ddag}${\small\it Yerevan Physics Institute\\ Alikhanian Br.
Str.
2, 0036 Yerevan, Armenia}\\
\medskip
{\small\tt manvel,ruehl@physik.uni-kl.de}
\vspace{2cm}
\begin{abstract}
The curvature of a higher spin potential as constructed in a previous article of the same authors  arXiv:0705.3528 is applied to the analysis of the linearized trace anomaly obtained from the quadratic part of the effective action for a conformally coupled scalar with  linearized interaction with  the external higher spin fields   arXiv:hep-th/0602067. The spin is restricted to four to profit from technical simplifications but without reducing the problem in principle. The issue includes the calculation of all Ricci tensors as multiple traces of the curvature, the derivation of all primary and secondary Bianchi identities, expressing all Ricci tensors as differential operators applied to the Fronsdal term, calculating the Weyl variation of these, and showing finally that Weyl variations of integrals over contracted squares of Ricci tensors can be used to eliminate the anomaly completely. This peculiarity is discussed in detail. As tools we use the formalism of bisymmetric tensor fields whose space is equipped with a local bilinear invariant form, the *-form.
\end{abstract}
\vspace{3cm}
{\it October 2007}
\end{center}
\newpage
\section{Introduction and motivation}
\quad
The conformal or trace anomaly \cite{anom} always unveils hidden connections of quantum field theory and the background geometry and topology. This should hold also in the case of conformal coupling of a scalar mode with the external higher spin field in a fixed $AdS_{4}$ background. The interest in this type of quantum field theory problem increased during the last years after the discovery of $AdS_{4}/CFT_{3}$ correspondence of the critical $O(N)$ sigma model and four dimensional higher spin gauge theory in anti-de-Sitter space  \cite{Kleb}. Investigation of this problem could also be important for
a deeper understanding of the geometrical and topological structure
of the linearized interaction of higher spin gauge fields \cite{Frons, Vasiliev}.
In this paper we apply the results of our previous article \cite{Curv} to explain the possible geometrical structure of the general formula for the trace anomaly in an external higher spin field in $AdS_{4}$ which is linearized in this field and was obtained in two other articles \cite{MR1, MR2} of ours.  More precisely the main motivation of this article is to find a geometrical interpretation of the formula for the general trace anomaly (107) of \cite{MR2}
with even spin. This anomalous trace can after some trivial algebraic manipulations be presented in the following elegant form
\begin{equation}\label{1.1}
    < J^{(s)\mu}_{\mu\mu_{3}\dots\mu_{s}}(z)>=
    \frac{\binom{2s}{s}[\Box-(s^{2}-1)]}{2^{s+4}\pi^{2}s(4s^{2}-1)}
    \prod^{s-2}_{m=0}\left(\Box+s-m(m+1)\right)h^{(s)\mu}_{\mu\mu_{3}\dots\mu_{s}}(z),
\end{equation}
where the current $J^{(s)}$ has spin $s$ and is on the classical level conserved and traceless  and has been constructed
from one scalar field with $s$ covariant derivatives in the $AdS_{d+1}$ space (see \cite{MR1,MR2} for details)\footnote{we use the same conventions as in \cite{MR2, Curv} for the  Euclidian $AdS_{d+1}$ metric and curvature\begin{eqnarray}
&&ds^{2}=g_{\mu \nu }(z)dz^{\mu }dz^{\nu
}=\frac{L^{2}}{(z^{0})^{2}}\delta _{\mu \nu }dz^{\mu }dz^{\nu
},\quad \sqrt{g}=\frac{L^{d+1}}{(z^{0})^{d+1}}\;,
\notag  \label{ads} \\
&&\left[ \nabla _{\mu },\,\nabla _{\nu }\right] V_{\lambda }^{\rho }=R_{\mu
\nu \lambda }^{\quad \,\,\sigma }V_{\sigma }^{\rho }-R_{\mu \nu \sigma
}^{\quad \,\,\rho }V_{\lambda }^{\sigma }\;,  \notag \\
&&R_{\mu \nu \lambda }^{\quad \,\,\rho
}=-\frac{1}{(z^{0})^{2}}\left( \delta _{\mu \lambda }\delta _{\nu
}^{\rho }-\delta _{\nu \lambda }\delta _{\mu }^{\rho }\right)
=-\frac{1}{L^{2}}\left( g_{\mu \lambda }(z)\delta _{\nu
}^{\rho }-g_{\nu \lambda }(z)\delta _{\mu }^{\rho }\right) \;,  \notag \\
&&R_{\mu \nu }=-\frac{d}{(z^{0})^{2}}\delta _{\mu \nu }=-\frac{d}{L^{2}}%
g_{\mu \nu }(z)\quad ,\quad R=-\frac{d(d+1)}{L^{2}}\;.  \notag
\end{eqnarray}}
 and $h^{(s)}_{\mu_{1}\mu_{2}\mu_{3}\dots\mu_{s}}(z)$ is a double traceless symmetric tensor higher spin external field as introduced by Fronsdal which we restricted to be transversal for simplicity ($\nabla^{\mu}h^{(s)}_{\mu\mu_{2}\mu_{3}\dots\mu_{s}}=0$).
Remember our result of \cite{MR2}, where the geometrical structure of this anomalous trace formula was obtained for general spin but analyzed only for the particular $s=2$ case. This
motivates the tasks for the next sections.

The recursive procedure of constructing a generalized curvature (and Christoffel symbols) for higher spin (HS) gauge fields in an $AdS_{d+1}$ background \cite{LV, Curv} allows us to perform the complete analysis for the first important case of the spin four gauge field (the most recent development in the vielbein formalism is considered in \cite{Hohm}). Evaluating the generalized curvature for $AdS_{4}$ and spin $s=4$ we perform then a calculation and classification of all its possible traces and all their possible Bianchi identities. This enables us to \emph{classify all gauge invariant local counterterms constructed from the contracted squares of generalized Riemann and Ricci tensors}.

In the subsequent section we will review our basic definitions, notations and formalism for
such calculations and present the formula for the curvature obtained in \cite{Curv} for general
spin $s$ which contains integer coefficients that are not all known explicitly. We present also some group theoretical arguments, in particular those clarifying the representation theoretic role of the deWit-Freedman form of the curvature \cite{DF}. In section 3 we present the generalized curvature for a spin four gauge field in explicit form and calculate all Ricci traces and formulate generalized Bianchi identities (a similar consideration for $s=3$ in the flat background see in \cite{Damour} ).
In section 4 we construct exploiting this formalism the full set of gauge invariant counterterms with independent generalized Weyl variation. The \emph{result} following from these calculations looks rather unexpected: \emph{Contrary to the $s=2$ case the number of local counterterms with independent Weyl variation for $s=4$ is in one-to-one correspondence with the numbers of independent structures in the trace anomaly formula and therefore we can cancel the linearized trace anomaly completely for $s=4$ using local counterterms}.
This on first glance strange result becomes more intuitive and evident when we remember that in the $s=2$ case that part of the anomaly which remains after cancellation is linearized in the $AdS_{4}$ background topological Euler density. So we could expect that after cancellation of
all regularization scheme dependent parts of the anomaly by the local counterterms, we would obtain the corresponding topological part for the higher spin case. But this cannot happen because in the $s\geq4$ case the traces of the corresponding currents are traceless tensors themselves. They can therefore not be connected with the metric independent topological objects which, even in the case of the generalization to higher spins, can only be scalar objects such as the Euler density.
Finally note that we do not consider the Weyl invariant part of the trace anomaly because this part of the trace starts from the second order in the external gauge field and therefore can be extracted only from the three point function (cubic part of the effective action). At this point the $s=2$ case is again special because, as it was explained in \cite{MR1}, from the two point function we can extract the
Weyl invariant divergent part of the effective action and then, because we know that our spin two gauge field is nothing but a fluctuation around our fixed $AdS_{4}$ background metric, we can restore the nonlinear divergent part and obtain the trace anomaly contribution. The major point of the considerations in \cite{MR1, MR2} was checking the anomalous coefficients in both cases (topological trace and Weyl invariant divergent part of effective action). The full agreement in both cases with textbook results \cite{BD} supported our confidence in the general formulas for the trace anomaly of one conformally coupled scalar mode in an external higher spin field (\ref{1.1}), and motivated us to develop and find some geometrical interpretation for these objects that were obtained just from a one loop Feynman diagram for the two point function \cite{MR1, MR2}. This will be done in this article and the emergence of higher spin geometrical objects in pure quantum expressions reflects the profound higher spin geometry nature of the linearized coupling of the scalar with the higher spin gauge field \cite{RM}.
\setcounter{equation}{0}
\section{Tensor fields, Young diagrams, and the \\
deWit-Freedman curvature}
Field theory on $AdS_{d+1}$ space has $O(d,1)$ as symmetry group. Local fields are sections through vector bundles
with base points $z\in AdS_{d+1}$  and representation spaces of $O(d,1)$ as fibres. If these representations are tensorial, they can be characterized, as has been shown long ago by H. Weyl, by Young diagrams which are in turn
ascribed to the unitary representations of the symmetry group $\bf{S}_{n}$. Here $n$ is the number of blocks in the diagram and the rank of the tensorial representation.

In a given basis of the vector representation of $O(d,1)$
\begin{equation}
v = \{v_{\mu} \mid 0\leq \mu \leq d \} ,
\label{2.1}
\end{equation}
we fill the Young diagram $Y$ with the labels $\mu$ in any order obtaining a Young "tableau".
\begin{equation}\label{2.2}
   Y=\tableau{{\ }&{\ }&{\ }&{\ }&{\ }&{\ }&{\ }&{\ }&{\ }\\{\ }&{\ }&{\ }&{\ }&{\ }&{\ }
   \\{\ }&{\ }&{\ }&{\ }&{\ }&{\ }\\{\ }&{\ }&{\ }}
\end{equation}
For each row $i$ of
length $n_{i}$ we have $n_{i}!$ permutations, we sum them. Then we multiply the result over all rows, getting an element
$\mathcal{S}_{Y}$ in the group algebra of the symmetric group $\bf{S}_{n}$, the "symmetrizer" of the diagram $Y$.
Now consider the columns $j$. For a fixed column we consider all permutations, multiply each with its signature and then sum them. Finally we multiply this result over all columns leading to an element $\mathcal{A}_{Y}$ of the group
algebra of $\bf{S}_{n}$, the "antisymmetrizer" of the diagram $Y$.
A tensor representation of rank $n$ projected on the tensor product of $n$ basis vectors (2.1)
has the symmetry of the diagram $Y$ if acting on the labels (filling them in the diagram from the upper left to the lower right, say) with $\mathcal{A}_{Y}
\mathcal{S}_{Y}$ leaves the tensor invariant, and we say that it has the symmetry of the
transpose diagram $Y_{t}$ if it is left invariant by application of first the antisymmetrizer and then the symmetrizer $\mathcal{S}_{Y} \mathcal{A}_{Y}$.

As an example consider first the higher spin potential $h^{(s)}(z), s \in \bf{N}$. It has the symmetry of a diagram $Y_{h}$ with one row of length $s$:
\begin{equation}\label{2.3}
    h^{(s)}_{\mu_{1}\dots\mu_{s}}(z)\Longrightarrow Y_{h}=\begin{tabular}{|c|cc|c|}
\hline
 $\mu_{1}$ & \,\,\,\,\,\,$\cdots$ &  & $\mu_{s}$ \\
\hline
\end{tabular}
\end{equation}
The Riemann curvature of $h^{(s)}$
(linearized in the potential $h^{(s)}$ in this article which does, however, not influence
the symmetry) has a diagram $Y_{R}$ with two rows of equal length $s$
\begin{equation}
R_{\mu_1\nu_1, \mu_2\nu_2,...\mu_{s}\nu_{s}}^{(s)}(z) \Longrightarrow Y_{R}=
\begin{tabular}{|c|c|cc|c|}
\hline
 $\mu_{1}$ & $\mu_{2}$ & \,\,\,\,\,$\cdots$ &  &$\mu_{s}$  \\
\hline
 $\nu_{1}$ & $\nu_{2}$ & \,\,\,\,\,$\cdots$ &  &$\nu_{s}$  \\
\hline
\end{tabular}
\label{2.4}
\end{equation}
But the deWit-Freedman curvature of $h^{(s)}$ \cite{DF} is defined to have the transpose diagram $Y_{Rt}$
as symmetry so that it can be obtained from the Riemann curvature by applying
the symmetrizer once again, and the Riemann curvature is recovered from the deWit-Freedman
one by application of the antisymmetrizer. Both forms are therefore different ways of organizing the same information.

Technically the most elegant way of handling symmetric tensors such as $h^{(s)}$ is by
contracting it with the $s'th$ tensorial power of a vector $a^{\mu}$ of the tangential space at
the base point $z$
\begin{equation}
h^{(s)}(z;a) = \sum_{\mu_{i}}(\prod_{i=1}^{s} a^{\mu_{i}})h^{(s)}_{\mu_1\mu_2...\mu_s}(z) .
\label{2.5}
\end{equation}
We obtain a homogeneous polynomial in the vector $a^{\mu}$ of degree $s$. The elegance of this method can e.g. inspected from the fact that requiring $h^{(s)}$ to be traceless, turns this polynomial into a Gegenbauer polynomial.
Applying the same method to the Riemann curvature we ought to contract it with the tensorial product of $s$ antisymmetric tensors of rank two \cite{Curv}. We shall not follow this idea in this article but use instead the deWit-Freedman curvature and contract it with the degree $s$
tensorial power of one tangential vector $a^{\mu}$ in the first row and with a similar tensorial power of another tangential vector $b^{\nu}$ in its second row. The effect of the additional symmetrizer is then explicit, but the action of the antisymmetrizer is hidden. Nevertheless it will become visible soon. The deWit-Freedman curvature is then written as
\begin{eqnarray}
\Gamma^{(s)}(z;a,b):\qquad
\Gamma^{(s)}(z;\lambda a,b) &=& \Gamma^{(s)}(z;a,\lambda b) \nonumber\\
&=& \lambda^{s}\Gamma^{(s)}(z;a,b)
\label{2.6}
\end{eqnarray}
We call such tensors depending on $a$ and $b$ "bisymmetric".

The group $O(d,1)$ possesses an invariant two-form, for its tensorial representations to be
irreducible they must be traceless. To achieve this one can extract traces that are obtained
as usual by cutting off two blocks from a row so that the result is still a Young diagram.
In the case of the symmetric tensors $h^{(s)}$ this is simple
\begin{equation}
Tr: h^{(s)}(z;a) \Longrightarrow Trh^{(s-2)}(z;a) = \frac{1}{s(s-1)}\Box_{a}h^{(s)}(z;a) .
\label{2.7}
\end{equation}
In the case of the deWit-Freedman curvature we can define $a$-traces, $b$-traces, and mixed traces. The $b$-trace is analogous to the trace of $h^{(s)}$
\begin{equation}
Tr_{b}: \Gamma^{(s)}(z;a,b) \Longrightarrow Tr_{b}\Gamma^{(s,s-2)}(z;a,b) =
\frac {1}{s(s-1)}\Box_{b}\Gamma^{(s)}(z;a,b) .
\label{2.8}
\end{equation}
The $a$-trace can be easily performed as follows. Due to our derivation in \cite{Curv} we know that
the deWit-Freedman curvature has certain properties which we quote here as propositions
(for easier quotation). First we have symmetry by exchange of $a$ and $b$
(Proposition 1):
\begin{equation}
\Gamma^{(s)}(z;a,b) = \Gamma^{(s)}(z;b,a) .
\label{2.9}
\end{equation}
Therefore the operation "$a$-trace" can be defined by (\ref{2.6}) with exchange of $a$ and $b$ at the end. The mixed trace is introduced by the operator
\begin{equation}
\frac{1}{s^{2}}(\partial_{a}\partial_{b}) ,
\label{2.10}
\end{equation}
and will be investigated in the subsequent section. Obviously the Riemann curvature has no mixed trace.

Directly connected with representation theory are manipulations involving other differentials with respect to $a$ and $b$, e.g.
\begin{eqnarray}
A_{b} = (a\partial_{b}) , \\
\label{2.11}
B_{a} = (b\partial_{a}) .
\label{2.12}
\end{eqnarray}
Then we can prove that (Proposition 2):
\begin{equation}
A_{b}\Gamma^{(s)}(z;a,b) = B_{a}\Gamma^{(s)}(z;a,b) = 0 .
\label{2.13}
\end{equation}
These "primary Bianchi identities" are manifestations of the hidden antisymmetry.

We remember that in \cite{Curv} we derived $\Gamma^{(s)}$ by three assumptions:
(1) the expansion in powers of the inverse $AdS$ radius
\begin{equation}
\Gamma^{(s)} = \sum_{k=0}^{s/2} L^{-2k} \Gamma^{(s)}_{k}\qquad (L:AdS \quad\textnormal{radius}) ;
\label{2.14}
\end{equation}
(2) the deWit-Freedman ansatz for the flat space term at $k=0$
\begin{equation}
\Gamma^{(s)}_{0}(z;a,b) = \sum_{l=0}^{s}\frac{(-1)^{l}}{l!} (a\nabla)^{l}(b\nabla)^{s-l}
B_{a}^{l}h^{(s)}(z;a) ;
\label{2.15}
\end{equation}
(3) the gauge invariance postulate: under an infinitesimal gauge transformation
\begin{equation}
\delta h^{(s)}(z;a) =(a\nabla)\epsilon^{(s-1)}(z;a) ,\quad \Box_{a}\epsilon^{(s-1)}(z;a)=0
\label{2.16}
\end{equation}
the deWit-Freedman curvature shall be invariant.
It turns out (Proposition 3) that assumption (3) can be replaced by either the first or the second primary Bianchi identity (\ref{2.13}). The result for the curvature is the same in all three cases. Remarks on the proof of this Proposition 3 can be found in Appendix B.

We want to close this section with (see \cite{Curv}) the remark
that the higher order terms in the expansion (\ref{2.14}) can be presented as
\begin{equation}
\Gamma^{(s)}_{k}(z;a,b) = \sum_{r_1r_2r_3} \sum_{l=l_{min}}^{l_{max}}\frac{(-1)^{l}}{l!}
A_{r_1r_2r_3}^{(l)}(a^2)^{r_1}(ab)^{r_2}(b^2)^{r_3}(a\nabla)^{l-l_{min}}(b\nabla)^{l_{max}-l}
B_{a}^{l}h^{(s)}(z;a) ,
\label{2.17}
\end{equation}
where the sum over the $r_{i}$ is restricted to $r_1+r_2+r_3 = k, l_{min} = 2r_1+r_2,
l_{max} = r_2+2r_3$, and the  coefficients $A_{r_1r_2r_3}^{(l)}$ are integers. They are given in \cite{Curv} for $k\in \{0,1,2\}$.

\setcounter{equation}{0}
\section{Alphabet of the  $s=4$ Curvature}

In this section we investigate all traces in the $s=4$ case.
The expression for $\Gamma^{(4)}(z;a,b)$ can be obtained from (\ref{2.15})-(\ref{2.17})
\begin{eqnarray}
   &&\Gamma^{(4)}(z;a,b) = \Gamma^{(4)}_{0}(z;a,b)+L^{-2}\Gamma^{(4)}_{1}(z;a,b)+L^{-4}\Gamma^{(4)}_{2}(z;a,b) , \label{3.1}\\
   &&\Gamma^{(4)}_{0}(z;a,b) = \sum_{l=0}^{4}\frac{(-1)^{l}}{l!} (a\nabla)^{l}(b\nabla)^{4-l}
B_{a}^{l}h^{(4)}(z;a) , \label{3.2}\\
   &&\Gamma^{(4)}_{1}(z;a,b) =\left\{a^{2}\sum_{l=2}^{4}\frac{(-1)^{l}}{l!}A_{100}^{(l)}
   (a\nabla)^{l-2}(b\nabla)^{4-l}   +(ab)\sum_{l=1}^{3}\frac{(-1)^{l}}{l!}A_{010}^{(l)}(a\nabla)^{l-1}(b\nabla)^{3-l}\right. \nonumber\\ && \qquad\qquad\qquad \left.+b^{2}\sum_{l=0}^{2}\frac{(-1)^{l}}{l!}A_{001}^{(l)} (a\nabla)^{l}(b\nabla)^{2-l} \right\}  B_{a}^{l}h^{(4)}(z;a) ,\label{3.3}\\
    &&\Gamma^{(4)}_{2}(z;a,b)= \left\{(4!)^{-1}a^{4}A^{(4)}_{200}B^{4}_{a}-(3!)^{-1}a^{2}(ab)A^{(3)}_{110}B^{3}_{a}
   +(2!)^{-1}a^{2}b^{2}A^{(2)}_{101}B^{2}_{a}\right.\nonumber\\
   &&\left. +(2!)^{-1}(ab)^{2}A^{(2)}_{020}B^{2}_{a}
   -(ab)b^{2}A^{(1)}_{011}B_{a}+b^{4}A^{(0)}_{002}\right\}h^{(4)}(z;a) ,\label{3.4}
\end{eqnarray}
and the following particular coefficients $A_{r_1r_2r_3}^{(l)}$ have been obtained in \cite{Curv}
\begin{eqnarray}
  A_{001}^{(l)} &=&-\binom{4-l}{3} , \quad l=0,1,2 ; \label{3.5}\\
  A_{010}^{(l)} &=& -l\binom{5-l}{2}, \quad l=1,2,3 ; \label{3.6}\\
  A_{100}^{(l)} &=& 2\binom{l}{3}-4\binom{l}{2}, \quad l=2,3,4 ;\label{3.7}\\
  A_{200}^{(4)} &=& 24 ;\quad A_{101}^{(2)} = A_{002}^{(0)} = 0 ;\label{3.8}\\
  A_{110}^{(3)} &=& 12 ; \quad A_{020}^{(2)}= 6 ; \quad A_{011}^{(1)} = 3 .\label{3.9}
\end{eqnarray}
Then we can start to investigate several traces of this curvature, so called generalized  Ricci tensors (see a similar classification for $s=3$ and flat background in \cite{Damour}). As it was mentioned in the previous section we can classify all independent traces using Young tableaus.
In this particular case it is the following expansion
\begin{eqnarray}
 \tableau{{\ a }&{\ }&{\ }&{\ }\\{\ b}&{\ }&{\ }&{\ }}&\Rightarrow&
 \tableau{{\ a}&{\ }&{\ }&{\ }\\{\ b}&{\ }}\quad\oplus\quad\tableau{{\ a}&{\ }\\{\ b}&{\ }} \quad\oplus\tableau{{\ a}&{\ }&{\ }&{\ }}\oplus\tableau{{\ a}&{\ }}\quad \oplus\,\emptyset\label{3.10}\\
 \Gamma^{(4)}(a,b)&\Rightarrow& \alpha(a,b)\quad\,\,\, \oplus\,\,\, \beta(a,b)\,\,\,\oplus\,\,\gamma(a)\,\quad\oplus\Delta(a)\quad\oplus\omega \label{3.11}
\end{eqnarray}
Comparing (\ref{3.10}) and (\ref{3.11}) we obtain homogeneity in $a$ and $b$ for these traces and write
\begin{eqnarray}
  \alpha(z;a,b) &=&Tr_{b}\Gamma^{(4)}(z;a,b) ;\label{3.12} \\
  \beta(z;a,b)&=&Tr_{a}\alpha(z;a,b)=Tr_{a} Tr_{b}\Gamma^{(4)}(z;a,b) ;\label{3.13}\\
  \gamma(z;a) &=& Tr_{b}\alpha(z;a,b)=Tr_{b}^{2}\Gamma^{(4)}(z;a,b)  ;\label{3.14}\\
  \Delta(z;a)&=&Tr_{a}\gamma(z;a)=Tr_{b}\beta(z;a,b)=Tr_{a}Tr_{b}^{2}
  \Gamma^{(4)}(z;a,b ;\label{3.15})\\
  \omega(z)&=&Tr_{a}\Delta(z;a)= Tr_{a}^{2}Tr_{b}^{2}\Gamma^{(4)}(z;a,b) .\label{3.16}
\end{eqnarray}
 All possible trace operations are epitomized in the following diagram for our "Ricci" alphabet
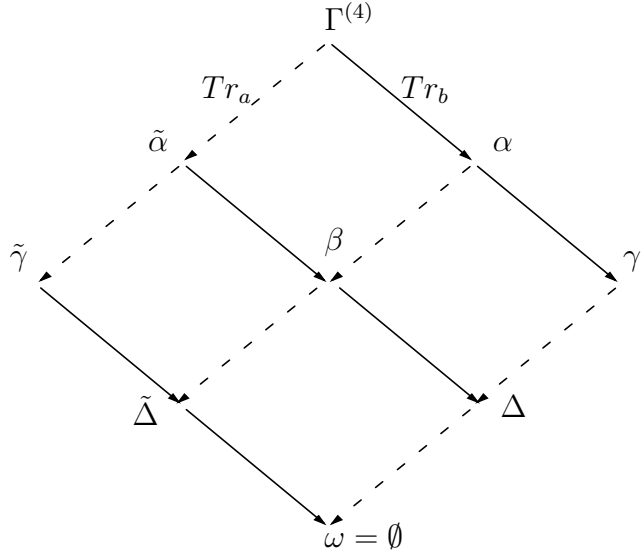
\begin{eqnarray}
\setlength{\unitlength}{0.252mm}
\begin{picture}(371,284)(100,-318)
        \allinethickness{0.252mm}\dashline{6}(264,-54)(192,-114)\special{sh 1}\path(192,-114)(197,-112)(196,-111)(195,-110)(192,-114) 
        \allinethickness{0.252mm}\path(268,-54)(340,-114)\special{sh
        1}\path(340,-114)(335,-112)(336,-111)(337,-110)(340,-114) 
        \allinethickness{0.252mm}\dashline{6}(416,-182)(344,-242)\special{sh 1}\path(344,-242)(349,-240)(348,-239)(347,-238)(344,-242) 
        \allinethickness{0.252mm}\path(192,-118)(264,-178)\special{sh 1}\path(264,-178)(259,-176)(260,-175)(261,-174)(264,-178) 
        \allinethickness{0.252mm}\dashline{6}(188,-118)(116,-178)\special{sh 1}\path(116,-178)(121,-176)(120,-175)(119,-174)(116,-178) 
        \allinethickness{0.252mm}\dashline{6}(260,-182)(188,-242)\special{sh 1}\path(188,-242)(193,-240)(192,-239)(191,-238)(188,-242) 
        \allinethickness{0.252mm}\path(116,-182)(188,-242)\special{sh 1}\path(188,-242)(183,-240)(184,-239)(185,-238)(188,-242) 
        \allinethickness{0.252mm}\path(344,-118)(416,-178)\special{sh 1}\path(416,-178)(411,-176)(412,-175)(413,-174)(416,-178) 
        \allinethickness{0.252mm}\path(272,-182)(344,-242)\special{sh 1}\path(344,-242)(339,-240)(340,-239)(341,-238)(344,-242) 
        \allinethickness{0.252mm}\dashline{6}(340,-118)(268,-178)\special{sh 1}\path(268,-178)(273,-176)(272,-175)(271,-174)(268,-178) 
        \allinethickness{0.252mm}\path(192,-246)(264,-306)\special{sh 1}\path(264,-306)(259,-304)(260,-303)(261,-302)(264,-306) 
        \allinethickness{0.252mm}\dashline{6}(340,-246)(268,-306)\special{sh 1}\path(268,-306)(273,-304)(272,-303)(271,-302)(268,-306) 
        \put(264,-46){\shortstack{$\Gamma^{(4)}$}} 
        \put(352,-110){\shortstack{$\alpha$}} 
        \put(172,-110){\shortstack{$\tilde{\alpha}$}} 
        \put(264,-162){\shortstack{$\beta$}} 
        \put(420,-170){\shortstack{$\gamma$}} 
        \put(100,-170){\shortstack{$\tilde{\gamma}$}} 
        \put(356,-250){\shortstack{$\Delta$}} 
        \put(164,-254){\shortstack{$\tilde{\Delta}$}} 
        \put(264,-318){\shortstack{$\omega=\emptyset$}} 
        \put(304,-82){\shortstack{$Tr_{b}$}} 
        \put(200,-82){\shortstack{$Tr_{a}$}} 
        
\end{picture}\nonumber\end{eqnarray}
which includes also the mirrored part obtained by $a\leftrightarrow b$ exchange
\begin{eqnarray}
  \tilde{\alpha}(a,b) &=& \alpha(b,a)=Tr_{a}\Gamma_{(4)}(a,b)\quad, \quad \textnormal{and so on.} \nonumber
\end{eqnarray}

To describe the mixed traces we turn to the primary Bianchi identities (\ref{2.13}). It is easy to see that any mixed trace can be expressed through the ordinary ones by taking a trace from one of the relations in (\ref{2.13}). For example
\begin{eqnarray}
  &&\Box_{b}A_{b}\Gamma^{(4)}(z;a,b)=0\quad\Rightarrow A_{b}\alpha(z;a,b)=0 ,\label{3.17}\\
  &&\Box_{a}A_{b}\Gamma^{(4)}(z;a,b)=0 \quad\Rightarrow (\partial_{a}\partial_{b})\Gamma^{(4)}(z;a,b)=-6A_{b}\tilde{\alpha}(z;a,b) , \label{3.18}\\
  && \Box_{a}A_{b}\alpha(z;,a,b)=0\quad\Rightarrow (\partial_{a}\partial_{b})\alpha(z;a,b)=-6A_{b}\beta(z;a,b) ,\label{3.19}\\
  &&\Box_{a}B_{a}\Gamma^{(4)}(z;a,b)=0\quad\Rightarrow B_{a}\tilde{\alpha}(z;a,b)=0 ,\label{3.20}\\
  &&\Box_{b}B_{a}\Gamma^{(4)}(z;a,b)=0\quad\Rightarrow (\partial_{a}\partial_{b})\Gamma^{(4)}(z;a,b)=-6B_{a}\alpha(z;a,b) ,\label{3.21}\\
  &&\Box^{2}_{b}B_{a}\Gamma^{(4)}(z;a,b)=0\quad\Rightarrow 2(\partial_{a}\partial_{b})\alpha(z;a,b)=-B_{a}\gamma(z;a,b) .\label{3.22}
  \end{eqnarray}
So we see immediately that (\ref{3.19}) and (\ref{3.22}) imply
\begin{equation}\label{3.23}
    B_{a}\gamma(z;a,b)=12A_{b}\beta(z;a,b) .
\end{equation}
The next interesting properties of the higher spin curvature and corresponding Ricci tensors are so called generalized secondary or differential Bianchi identities. We can formulate  these identities  in our notation in the following compressed  form ($[\dots]$ is antisymmetrization )
\begin{equation}\label{3.24}
    \frac{\partial}{\partial a^{[\mu}}\frac{\partial}{\partial b^{\nu}}\nabla_{\lambda]}\Gamma^{(4)}(z;a,b)= \Delta^{B}_{\mu\nu\lambda}\Gamma^{(4)}(z;a,b)=0 .
\end{equation}
Then as before we can contract (\ref{3.23}) with some combination of $a, b$, and $\partial_{b}, \partial_{a}$ and get identities for our alphabet of traces or Ricci tensors.
The most useful one we obtain from
\begin{equation}\label{3.25}
    a^{\mu}b^{\nu}\partial^{\lambda}_{b}\Delta^{B}_{\mu\nu\lambda}\Gamma^{(4)}(z;a,b)=0.
\end{equation}
After some algebra and using (\ref{2.13}) and (\ref{3.17})-(\ref{3.22}) we obtain
the following relation between the divergence of $\alpha$ and gradients of $\gamma$
\begin{equation}\label{3.26}
    (\nabla\partial_{b})\alpha(z;a,b)=2(b\nabla)
    \gamma(z;a)-\frac{1}{2}(a\nabla)B_{a}\gamma(z;a) .
\end{equation}
Moreover taking different including mixed traces and using again primary identities we can derive
\begin{eqnarray}
  (\nabla\partial_{b})\beta(z;a,b)&=&\frac{7}{4}(b\nabla)\Delta(z;a)
  -\frac{3}{4}(a\nabla)B_{a}\Delta(z;a) ,\label{3.27}\\
  (\nabla\partial_{a})\gamma(z;a)&=& 3(a\nabla)\Delta(z;a) .\label{3.28}
\end{eqnarray}
The properties of the last scalar "Ricci" $\omega(z)$ we will be described in the next section.

Finally we would like to present one more gauge invariant object that has not been listed before. It is the so-called "Fronsdal term", a second order differential operator (Fronsdal operator) which is applied to the higher spin gauge field and which, set equal zero, defines the free field equation of motion (from now on we put $AdS$ radius $L=1$).
\begin{eqnarray}
\mathcal{F}(h^{(s)}(z;a))&=&\Box h^{(s)}(z;a)-(a\nabla
)(\nabla\partial_a)h^{(s
)}(z;a)+\frac{1}{2}(a\nabla
)^{2}\Box _{a}h^{(s )}(z;a)\quad   \nonumber \\
&-&\left( s ^{2}+s(d-5)-2(d-2)\right) h^{(s)}(z;a)-a^{2}\Box
_{a}h^{(s)}(z;a) .  \label{3.29}
\end{eqnarray}
The normalization of the Fronsdal term is ad hoc obviously. It satisfies the following "Bianchi" identity
\begin{equation}\label{3.30}
    (\nabla\partial_{a})\mathcal{F}=\frac{s(s-1)}{2}(a\nabla)Tr_{a}\mathcal{F} .
\end{equation}
Inserting in these expressions $s=4$ and $d=3$ ($AdS_{4}$) we obtain the following forms of the Fronsdal term itself and it's trace
\begin{eqnarray}
  &&\mathcal{F}(h^{(4)}) = [\Box-8] h^{(4)}
  -(a\nabla)(\nabla\partial_{a})h^{(4)}+\left(6(a\nabla)^{2}-12a^{2}\right)h^{(4)} ,\label{3.31} \\
 && Tr_{a}\mathcal{F}(h^{(4)}) = 2[\Box-15]Tr_{a}h^{(4)}-\frac{1}{6}(\nabla\partial_{a})^{2}h^{(4)}
  +(a\nabla)(\nabla\partial_a)Tr_{a}h^{(4)} \label{3.32} , \\
  &&(\nabla\partial_{a})\mathcal{F}=6(a\nabla)Tr_{a}\mathcal{F} . \label{3.33}
   \end{eqnarray}

\setcounter{equation}{0}
\section{Local Counterterms and Trivialization of the\\  Trace Anomaly for $s=4$}

In this section we are going to calculate directly the complete hierarchy of "Ricci" tensors for the deWitt-Freedman curvature at $s=4$ (\ref{3.1}) and express them through the Fronsdal terms (\ref{3.30}), (\ref{3.31}). The idea why all members of the alphabet, each of which contains derivatives of the fourth order of $h^{(4)}$, should be encoded with a single second order operator
is the following: after taking even a first trace we obtain at least one Laplacian. Then we can rearrange $AdS$ covariant derivatives shifting the Laplacian to the front of $h^{(4)}(z;a)$ (remember that the whole curvature is linear in $h^{(4)}$). But the gauge invariant extension of the Laplacian is unique, it is just the Fronsdal operator. For a flat space background this statement is absolutely clear and can be proven easily. The problem in $AdS$ space is that there are several $O(L^{-2})$ and $O(L^{-4})$ expressions coming from the curvature itself and from the commutation of covariant derivatives. They include not only terms proportional to $a^{2}$ and $b^{2}$ but also contact terms with $(ab)$. This could produce in principle some traceless combination and other projectors existing only in this bisymmetric space of tensors such as $\Gamma^{(4)}(z;a,b)$. So to be sure that after factorization of the traces in the second order differential operator times $\mathcal{F}(h^{(4)})$ there are no residual terms we performed direct calculations of all the traces of the curvature (\ref{3.1}) with all $O(L^{-2})$ and $O(L^{-4})$ terms and observed exact factorization of the Fronsdal term in all orders of $AdS$ radius.
During these tedious calculations we used some interesting relations for commutators of covariant derivatives acting in the bisymmetric space of tensors $T^{(n)(m)}(z;a,b)$.
To avoid complicated and long formulas in the body of this article we will present here only  the results  and refer the interested reader for the details on commutations to the  Appendix A.

So finally we obtain the following expressions for our "alphabet" of Ricci tensors
\begin{eqnarray}
   \alpha(z;a,b)&=& \left[(b\nabla)^{2}-\frac{1}{2}(a\nabla)(b\nabla)B_{a} + \frac{1}{12}(a\nabla)^{2}B^{2}_{a}-\frac{1}{3}a^{2}B^{2}_{a}+(ab)B_{a}\right]\mathcal{F} , \label{4.1}\\
\beta(z;a,b)&=& \frac{1}{72}[\Box+4]B^{2}_{a}\mathcal{F}+ \left[\frac{1}{2}(b\nabla)^{2}-\frac{2}{3}(a\nabla)(b\nabla)B_{a}
 \right.\nonumber\\ && \qquad\qquad\qquad\qquad\left.+\frac{1}{4}(a\nabla)^{2}B^{2}_{a}-\frac{2}{3}a^{2}B^{2}_{a}
 +\frac{1}{3}(ab)B_{a}\right]Tr_{a}\mathcal{F} ,\label{4.2}\\
 \gamma(z;a)&=&[\Box+4]\mathcal{F}-2\left[(a\nabla)^{2}
 +2a^{2}\right]Tr_{a}\mathcal{F} ,\label{4.3}\\
 \Delta(z;a)&=&\frac{2}{3}[\Box+2
 -(a\nabla)(\nabla\partial_{a})]Tr_{a}\mathcal{F} ,\label{4.4}\\
 \omega(z)&=&-\frac{2}{3}(\nabla\partial_{a})^{2}Tr_{a}\mathcal{F} .\label{4.5}
\end{eqnarray}
Then after a short calculation using (\ref{3.33}) and commutation relations from Appendix A we see that
\begin{equation}\label{4.6}
    (\nabla\partial_{a})Tr_{a}\mathcal{F}=-\frac{1}{6}(\nabla\partial_{a})^{3}h^{(4)}
    +\left[(3\Box-43)(\nabla\partial_{a})+(a\nabla)
    (\nabla\partial_{a})^{2}\right]Tr_{a}h^{(4)} .
\end{equation}
This and (\ref{4.5}) show that contrary to the spin two case the scalar Ricci tensor $\omega$ is completely longitudinal and does not carry physical information. The analog of the gravitational Ricci scalar here is the second rank symmetric Ricci tensor $\Delta$ and we see below that this tensor will play the most important role in the investigation of the anomaly (\ref{1.1}).

Now we start to analyse the structure of (\ref{1.1}) for the case $s=4$
\begin{eqnarray}
    < Tr_{a} J^{(4)}(z;a)>&=&
    \frac{5}{2^{9}9\pi^{2}}[\Box-15][\Box+4][\Box+2][\Box-2]Tr_{a}h^{(4)}(z;a),\label{4.7}\\
    (\nabla\partial_{a})h^{(4)}(z;a)&=&(\nabla\partial_{a})Tr_{a}h^{(4)}(z;a)=0 .\label{4.8}
\end{eqnarray}
Comparing with (\ref{4.4}) and taking into account (\ref{4.6}) we obtain immediately the following nice relation for a transversal external higher spin field
(from now on we will always put the longitudinal part of $h^{(4)}$ to zero at the end of calculations)
\begin{equation}\label{4.9}
     < Tr_{a} J^{(4)}(z;a)>=
    \frac{5}{2^{11}3\pi^{2}}[\Box+4][\Box-2]\Delta(z;a) .
\end{equation}
Then we start to classify the possible local counterterms.

It is clear that these should be quadratic in the curvature and Ricci tensors
\begin{equation}\label{4.10}
    K^{c^{(i)}}=\frac{1}{2}\int\sqrt{g}d^{4}z c^{(i)}(z;a,b)*c^{(i)}(z;a,b) ,
\end{equation}
where
\begin{equation}\label{4.11}
    c^{(i)}(z;a,b)\in \left\{\Gamma^{(4)}(z;a,b), \alpha(z;a,b), \beta(z;a,b), \gamma(z;a,b), \Delta(z;a,b)\right\} ,
\end{equation}
and we introduced the notation $*$ for a contraction in the bisymmetric space of indices $(a,b)$
\begin{eqnarray}
  *&=&\frac{1}{(4!)^{2}} \prod^{4}_{i,j=1}\overleftarrow{\partial}^{\mu_{i}}_{a}\overrightarrow{\partial}_{\mu_{i}}^{a}
  \overleftarrow{\partial}^{\mu_{j}}_{b}\overrightarrow{\partial}_{\mu_{j}}^{b} \label{4.12}
\end{eqnarray}
Then we see that operators $A_{b}, a^{2}, b^{2}$ are dual (or adjoint) to $B_{a},\Box_{a},\Box_{b}$ with respect to the "star" product (\ref{4.12})
\begin{eqnarray}
    A_{b}f(a,b)*g(a,b)&=& f(a,b)*B_{a}g(a,b)\label{4.13}\\
    {a^{2}\atop b^{2}}f(a,b)*g(a,b)&=&f(a,b)*{\Box_{a}\atop\Box_{b}}g(a,b) . \label{4.14}
\end{eqnarray}
In the same fashion gradients and divergences are dual with respect to the full scalar product in the space $(z,a,b)$
\begin{eqnarray}
  \int\sqrt{g}d^{4}z {(a\nabla)\atop(b\nabla)}f(z;a,b)*g(z;a,b) &=& -\int\sqrt{g}d^{4}z f(z;a,b)*{(\nabla\partial_{a})\atop(\nabla\partial_{b})}g(z;a,b) .\nonumber\\ \label{4.15}
  \end{eqnarray}
\quad Now we collected all the tools including Appendix A for the study of the behaviour of the local counterterms (\ref{4.10}) under generalized Weyl transformations with a traceless tensor parameter defined in \cite{Damour} and for spin $s=4$ investigated in \cite{RM}
\begin{eqnarray}\label{4.16}
    \delta h^{(4)}(z;a)=a^{2}\sigma^{(2)}(z;a), \quad \Box_{a}\sigma^{(2)}(z;a)=0 .
\end{eqnarray}
It is a direct generalization of the usual Weyl transformation of the linearized gravitational field and our trace anomaly (\ref{1.1}) is the variation of the quantum effective action with respect to this generalized Weyl transformation. But we note immediately that there is one evident constraint on the variations of $\delta K^{c^{(i)}}$: It is the existence of the generalized Weyl tensor. Indeed we can extract all traces from $\Gamma^{4}$ and obtain a completely traceless and conformally invariant (in means of (\ref{4.16})) curvature $W^{(4)}(z;a,b)$. Therefore there is a fixed combination of our counterterms (\ref{4.10})
\begin{equation}\label{4.17}
    \int\sqrt{g}d^{(4)}z W^{4}(z;a,b)*W^{(4)}(z;a,b) ,
\end{equation}
which is Weyl invariant. So we see that not all our counterterms do have independent Weyl variations and we can drop one of the tensors from the set (\ref{4.11}), say $\Gamma^{(4)}$.
Then due to the factorization (\ref{4.1})-(\ref{4.4}) we  see that the Weyl variation of the  alphabet of curvatures $c^{(i)}\in\{\alpha, \beta, \gamma, \Delta\}$ are always of the following form
\begin{eqnarray}
    \delta c^{(i)}&=&\hat{M}((\nabla\partial_{a}), \Box, (a\nabla),B_{a},\dots)\delta \mathcal{F}(z,a)\nonumber\\&+&\hat{N}((\nabla\partial_{a}), \Box, (a\nabla),B_{a}\dots)\delta Tr_{a}\mathcal{F}(z,a) ,\label{4.18}
\end{eqnarray}
where the differential operators $\hat{M}((\nabla\partial_{a}), \Box, (a\nabla),B_{a},\dots)$ and $\hat{N}((\nabla\partial_{a}), \Box, (a\nabla),B_{a},\dots)$ can be easily inspected from (\ref{4.1})-(\ref{4.4}). So we have only to calculate the Weyl variation of  $\mathcal{F}$ and $Tr_{a}\mathcal{F}$
\begin{eqnarray}
  \delta \mathcal{F}(z,a) &=& 6(a\nabla)^{2}\sigma^{(2)}+a^{2}\left[\Box-(a\nabla)
  (\nabla\partial_{a})-22\right]\sigma^{(2)} ,\label{4.19}\\
   \delta Tr_{a}\mathcal{F}(z,a) &=& \frac{7}{3}(\Box-16)\sigma^{(2)}+\frac{2}{3}(a\nabla)(\nabla\partial_{a})
   \sigma^{(2)}-\frac{1}{6}a^{2}(\nabla\partial_{a})^{2}\sigma^{(2)} .\label{4.20}
\end{eqnarray}
Then performing the variation of (\ref{4.10}) and using duality (\ref{4.13}), (\ref{4.14}) we obtain
\begin{eqnarray}
    \delta K^{c^{(i)}}&=&\int\sqrt{g}d^{4}z \delta \mathcal{F}(z,a)*\hat{M}(-(a\nabla), \Box,-(\nabla\partial_{a}), A_{b},\dots)c^{(i)}(z;a,b)\nonumber\\
    &+& \int\sqrt{g}d^{4}z \delta Tr_{a}\mathcal{F}(z,a)*\hat{N}(-(a\nabla), \Box,-(\nabla\partial_{a}), A_{b},\dots)c^{(i)}(z;a,b) .\label{4.21}
\end{eqnarray}
Now we can insert here (\ref{4.19}) and (\ref{4.20}) and working in a similar fashion as above using duality, Bianchi identities, and several numbers of the commutation relations from Appendix A, we arrive at the following list of variations putting the longitudinal part of the gauge field to zero at the end
\begin{eqnarray}
  \frac{1}{65}\frac{\delta K^{\alpha}}{\delta \sigma^{(2)}(z;a)} &=& \left[\Box^{2}-12\Box-\frac{592}{13}\right]\Delta(z;a) ;\label{4.22}\\
  \frac{9}{59}\frac{\delta K^{\beta}}{\delta \sigma^{(2)}(z;a)}&=& \left[\Box^{2}-13\Box-\frac{2820}{59}\right]\Delta(z;a) ;\label{4.23}\\
  \frac{1}{16}\frac{\delta K^{\gamma}}{\delta \sigma^{(2)}(z;a)}&=& \left[\Box^{2}-13\Box-30\right]\Delta(z;a) ;\label{4.24}\\
   \frac{9}{28}\frac{\delta K^{\Delta}}{\delta \sigma^{(2)}(z;a)}&=& \left[\Box^{2}-14\Box-32\right]\Delta(z;a) .\label{4.25}
\end{eqnarray}
Note that the same $\Delta(z;a)$ arose in all variations according to the Bianchi identities.

We see immediately that from these four variations only three are linearly independent enabling us to drop one more counterterm say $K^{\alpha}$. This happens because in four dimensions there is another Weyl invariant combination of our counterterms. Indeed we can contract two generalized curvatures (\ref{2.6}) in dimension four, using four totally antisymmetric Levi-Civita tensors, and integrate. This integral will be again a combination of our counterterms and on the other hand an integral over a total derivative due to Bianchi identities and is therefore Weyl invariant. So in this respect we notice an analogy with the Euler density for the graviton. Thus we got only three counterterms
$K^{\Delta}, K^{\gamma}, K^{\beta}$ with independent generalized Weyl variations (\ref{4.23})-(\ref{4.25}) and using a linear combination of these we can cancel any object with the structure
\begin{equation}\label{4.26}
    \left[\Box^{2}+m_{1}\Box+m_{2}\right]\Delta(z;a),\quad\quad m_{1}, m_{2}\quad \textnormal{is arbitrary numbers}.
\end{equation}
But our trace anomaly (\ref{4.9}) has a structure just of this type
\begin{equation}\label{4.27}
    \frac{2^{11}3\pi^{2}}{5} < Tr_{a} J^{(4)}(z;a)>=
    [\Box^{2}+2\Box-8]\Delta(z;a) ,
\end{equation}
and we see that we can cancel the trace anomaly of a conformally coupled scalar in the spin four external tensor background completely. So we obtained the result promised in the
introduction.

At the end of this section note that the situation in the case of spin two is completely different. Remembering our consideration in \cite{MR2} for the gravitational case we had a corresponding structure for the anomaly as $[\Box+m_{1}]R_{lin}$, where $R_{lin}$ was the  linearized Ricci scalar ($s=2$ analog of $\Delta$) and only one independent counterterm  with Weyl variation $\Box R_{lin}$. Thus after a possible cancelation the remaining term could be identified with the linearized Euler density carrying the proper coefficient in front. The reason of such a dramatic difference is the existence of a nontrivial background geometry for the graviton. As a result the local counterterm is  $R_{nonlin}^{2}$, where $R_{nonlin}=R_{AdS}+R_{linear}+O([h^{(2)}]^{2})+\dots$
with a variation completely different from the variation of $R_{lin}^{2}$. In the higher spin case where we have no background, all our linearized counterterms start from the squares of the linearized curvature and Ricci tensors.

\section{Conclusion}
\setcounter{equation}{0}
We have proved that the anomaly obtained from the quadratic part of the effective action (two-point function of the  higher spin currents constructed from one scalar field and $s$ derivatives) which is linear in the higher spin gauge field can be cancelled completely by renormalization terms formed from integrals over contracted Ricci tensors. This proof is given here only for spin four fields.
This has the advantage to simplify the explicit and often very tedious calculations, but all
general aspects of the problem remain unchanged and the tools applied are all generic.
The algorithm consists of several steps which can be formulated in such a fashion that
computer programs for the general case are possible. This is also true for the curvature
calculated in \cite{Curv}: it is complete for spin four but the steps for the general case are mathematically well prepared.
A problem of quite another quality is to generalize our approach to the second order part of anomalies connected with an investigation of the three-point function. No such three-point function has ever been calculated, but the analysis of an expected anomaly is certainly of great interest.

\subsection*{Acknowledgements}
\quad This work was supported by Alexander von Humboldt Foundation under 3.4-Fokoop-ARM/1059429.

\section*{Appendix A: Commutators}
\setcounter{equation}{0}
\renewcommand{\theequation}{A.\arabic{equation}}
The most important formula for performing all calculations in bisymmetric tensor space
\begin{equation}\label{A1}
    T^{(n)(m)}(z;a,b)=T_{\mu_{1}\dots\mu_{n},\nu_{1}\dots\nu_{n}}a^{\mu_{1}}\dots a^{\mu_{n}}b^{\nu_{1}}\dots b^{\nu_{m}} ,
\end{equation}
is the commutator of covariant derivatives:
\begin{equation}\label{A2}
    \left[\nabla_{\mu},\nabla_{\nu}\right]=a_{[\nu}\partial^{a}_{\mu]}
    +b_{[\nu}\partial^{b}_{\mu]} .
\end{equation}
This is in agreement with our conventions for the curvature of $AdS_{d+1}$ from the footnote in the introduction.
Then multiplying and contracting with the different combinations of $a,b$ and their derivatives we can obtain the following general formulas
\begin{eqnarray}
&&\left[(a\nabla),(b\nabla)\right]= b^{2}A_{b}-a^{2}B_{a}+(ab)[(a\partial_{a})-(b\partial_{b})] ;\label{A.3}\\
&& \left[(\nabla\partial_{a}),(a\nabla)\right] = \Box +a^{2}\Box_{a}-B_{a}A_{b}+(b\partial_{b})-(d-1)(a\partial_{a})
  -(a\partial_{a})^{2}+(ab)(\partial_{a}\partial_{b}) ;\quad\quad\label{A.4}\\
&&\left[(\nabla\partial_{a}),(b\nabla)\right] = (ab)\Box_{a} -B_{a}[(a\partial_{a})+(b\partial_{b})]-(d-1)B_{a}
 +b^{2}(\partial_{a}\partial_{b}) .\label{A.5}
\end{eqnarray}
The relation for $\left[(\nabla\partial_{b}),(b\nabla)\right]$ and $\left[(\nabla\partial_{b}),(a\nabla)\right]$ can be obtained from (\ref{A.4}) and (\ref{A.5})
with exchange of $a\leftrightarrow b$ and $A_{b}\leftrightarrow B_{a}$.

At the end we present all important commutation relations working in the space of symmetric rank $n$ tensors
\begin{eqnarray}
  &&[(\nabla\partial_{a}), \Box]f^{(n)}(z,a)= \left[2(a\nabla)\Box_{a}-(d+2n-2)(\nabla\partial_{a})\right]f^{(n)}(z,a) ;\label{A.6}\\
  &&[(\nabla\partial_{a}), (a\nabla)]f^{(n)}(z,a)= \Box f^{(n)}(z,a)
 + [\nabla_{\mu}, (a\nabla)]\partial_{a}^{\mu}f^{(n)}(z,a) ;\label{A.7}\\
&& [\nabla_{\mu}, (a\nabla)]\partial_{a}^{\mu}f^{(n)}(z,a)= \left[a^{2}\Box_{a}-n(d+n-1)\right]f^{(n)}(z,a) ;\label{A.8}\\
&&[\Box ,(a\nabla)]f^{(n)}(z,a)= \left[2a^{2}(\nabla\partial_{a})-(d+2n)(a\nabla)\right]f^{(n)}(z,a) ;\label{A.9}\\
&&\Box_{a}\left[a^{2}f^{(n)}(z,a)\right] = 2(d+2n+1)f^{(n)}(z,a)+a^{2}\Box_{a}f^{(n)}(z,a) .\label{A.10}
\end{eqnarray}

\section*{Appendix B: Remarks on Proposition 3}
\setcounter{equation}{0}
\renewcommand{\theequation}{B.\arabic{equation}}
The construction of the curvature in \cite{Curv} was achieved for all spins but only for
the first three levels $k\leq2$ due to the enormous complication in the difference equations
\cite{Curv}, (\ref{4.1}). The situation we meet when we use the Bianchi identity
(\ref{2.13}) is only slightly different. Instead of the system based on the gauge invariance postulate
\begin{equation}
A_{r_1r_2r_3}^{(l+1)} - A_{r_1r_2r_3}^{(l)} = R_{r_1r_2r_3}(l) ,
\label{B.1}
\end{equation}
we obtain for the postulate based on the primary Bianchi identity involving $A_{b}$
\begin{equation}
\frac{A_{r_1r_2r_3}^{(l+1)}}{{l+1 \choose 2r_1+r_2}} - \frac{A_{r_1r_2r_3}^{(l)}}{{l\choose 2r_1+r_2}} = R'_{r_1r_2r_3}(l) .
\label{B.2}
\end{equation}
Solving the second system of difference equations explicitly we obtain the same solutions as from the first system. The second system has one formal advantage: There exist in each case (depending on the $r_{i}$) only one boundary value which fixes the solution uniquely. In the first system we have two boundary conditions whenever $2r_1+r_2 > 0$ and one has to prove separately that this does not lead to an obstruction of the system. It does not, as we
learnt in \cite{Curv}.

\end{document}